\documentclass[]{elsarticle}
\usepackage[margin=1in]{geometry}
\usepackage{lineno,hyperref}
\modulolinenumbers[5]
\usepackage{multirow}
\usepackage{array}
\usepackage{amsmath}
\usepackage{bm}
\newcolumntype{P}[1]{>{\centering\arraybackslash}p{#1}}
\newcolumntype{M}[1]{>{\centering\arraybackslash}m{#1}}
\journal{Elsevier}
\usepackage{pifont}
\newcommand{\cmark}{\ding{51}}%
\newcommand{\xmark}{\ding{55}}%

\makeatletter
\def\ps@pprintTitle{%
 \let\@oddhead\@empty
 \let\@evenhead\@empty
 \let\@oddfoot\@empty
 \let\@evenfoot\@empty}
\makeatother

\begin{document}

\begin{frontmatter}

\title{ISLES'24: Final Infarct Prediction with Multimodal Imaging and Clinical Data. Where Do We Stand?}

\author[1]{Ezequiel de la Rosa\corref{cor1}}
\author[2]{Ruisheng Su}
\author[3,4]{Mauricio Reyes}
\author[9]{Evamaria O. Riedel}
\author[7,26]{Hakim Baazaoui}
\author[5,6]{Roland Wiest}
\author[1,9,12]{Florian Kofler}
\author[1]{Kaiyuan Yang}
\author[10]{David Robben}
\author[14]{Mahsa Mojtahedi}
\author[14]{Laura van Poppel}
\author[14]{Lucas de Vries}
\author[15,16]{Anthony Winder}
\author[15,16]{Kimberly Amador}
\author[15,16,17]{Nils D. Forkert}
\author[18]{Gyeongyeon Hwang}
\author[18]{Jiwoo Song}
\author[18]{Dohyun Kim}
\author[15]{Eneko Uruñuela}
\author[15]{Annabella Bregazzi}
\author[15]{Matthias Wilms}
\author[19]{Hyun Yang}
\author[19]{Jin Tae Kwak}
\author[19]{Sumin Jung}
\author[20]{Luan Matheus Trindade Dalmazo}
\author[20]{Kumaradevan Punithakumar}
\author[21]{Moona Mazher}
\author[22]{Abdul Qayyum}
\author[22]{Steven Niederer}
\author[23]{Jacob Idoko}
\author[23]{Mariana Bento}
\author[23]{Gouri Ginde}
\author[24]{Tianyi Ren}
\author[24]{Juampablo Heras Rivera}
\author[24]{Mehmet Kurt}
\author[25]{Carole Frindel}
\author[7,8]{Susanne Wegener}
\author[9,11]{Jan S. Kirschke}

\author[11,13]{Benedikt Wiestler\textsuperscript{$\dagger$}}
\author[1]{Bjoern Menze\textsuperscript{$\dagger$}}

\cortext[cor1]{Corresponding author: ezequiel.delarosa@uzh.ch}
\cortext[cor2]{\textsuperscript{$\dagger$}Equal contribution}

\address[1]{Department of Quantitative Biomedicine, University of Zurich, Zurich, Switzerland.}
\address[2]{Department of Biomedical Engineering, Eindhoven University of Technology, Eindhoven, the Netherlands.}
\address[3]{ARTORG Center for Biomedical Research, University of Bern, Bern, Switzerland.}
\address[4]{Department of Radiation Oncology, University Hospital Bern, University of Bern.}
\address[5]{Support Center of Advanced Neuroimaging (SCAN), University Institute of Diagnostic and Interventional Neuroradiology, Inselspital, Bern, Switzerland.}
\address[6]{University Institute of Diagnostic and Interventional Neuroradiology, University Hospital Bern, Inselspital, University of Bern, Bern, Switzerland.}
\address[7]{Department of Neurology, University Hospital of Zurich, Zurich, Switzerland.}
\address[8]{University of Zurich, Zurich, Switzerland.}
\address[9]{Department of Diagnostic and Interventional Neuroradiology, School of Medicine and Health, TUM Klinikum, Technical University of Munich, Germany.}
\address[10]{icometrix, Leuven, Belgium}
\address[11]{TranslaTUM, Center for Translational Cancer Research, Technical University of Munich, Germany.}
\address[12]{Helmholtz AI, Helmholtz Munich, Germany.}
\address[13]{AI for Image-Guided Diagnosis and Therapy, School of Medicine and Health, Technical University of Munich, Munich, Germany.}
\address[14]{Department of Biomedical Engineering and Physics, Amsterdam UMC location University of Amsterdam, Amsterdam, the Netherlands}
\address[15]{Department of Radiology, University of Calgary, Calgary, Canada}
\address[16]{Hotchkiss Brain Institute, University of Calgary, Calgary, Canada}
\address[17]{Alberta Children’s Hospital Research Institute, University of Calgary, Calgary, Canada}
\address[18]{Heuron Co., Ltd., Seoul, South Korea}
\address[19]{School of Electrical Engineering, Korea University, Seoul, Korea}
\address[20]{University of Alberta, Alberta, Canada}
\address[21]{Hawkes Institute, Department of Computer Science, University College London, London, United Kingdom}
\address[22]{National Heart and Lung Institute, Faculty of Medicine, Imperial College London, London, United Kingdom}
\address[23]{University of Calgary, Canada}
\address[24]{University of Washington, Washington, United States}
\address[25]{CREATIS, Universite Claude Bernard Lyon 1, INSA Lyon, UMR CNRS 5220, Inserm U1294, Villeurbanne,
France}
\address[26]{Department of Neurology, University Hospital Zurich, Zurich, Switzerland}

\begin{abstract}
Accurate estimation of brain infarction (i.e., irreversibly damaged tissue) is critical for guiding treatment decisions in acute ischemic stroke. Reliable infarct prediction informs key clinical interventions, including the need for patient transfer to comprehensive stroke centers, the potential benefit of additional reperfusion attempts during mechanical thrombectomy, decisions regarding secondary neuroprotective treatments, and, ultimately, prognosis of clinical outcomes. This work introduces the Ischemic Stroke Lesion Segmentation (ISLES) 2024 challenge, which focuses on the prediction of final infarct volumes from pre-interventional acute stroke imaging and clinical data. ISLES'24 provides a comprehensive, multimodal setting where participants can leverage all clinically and practically available data—including full acute CT imaging, sub-acute follow-up MRI, and structured clinical information—across a train set of $N = 150$ cases. On the hidden test set ($N = 98$), the top-performing model, a multimodal nnU-Net-based architecture, achieved a Dice score of $0.285 \pm 0.213$ and an absolute volume difference of $21.2 \pm 37.2$~mL, underlining the significant challenges posed by this task and the need for further advances in multimodal learning. This work makes two primary contributions: first, we establish a standardized, clinically realistic benchmark for post-treatment infarct prediction, enabling systematic evaluation of multimodal algorithmic strategies on a longitudinal stroke dataset; second, we analyze current methodological limitations and outline key research directions to guide the development of next-generation infarct prediction models.
\end{abstract}

\begin{keyword}
\texttt Ischemic stroke \sep Lesion segmentation \sep Final infarct \sep Multimodal data 
\end{keyword}

\end{frontmatter}

\section{Introduction}

Accurate segmentation of ischemic stroke lesions is crucial for guiding treatment decisions during acute stages (e.g., determining eligibility for thrombectomy), evaluating outcomes, clinical follow-up, and optimizing therapeutic strategies in sub-acute and chronic stages. The Ischemic Stroke Lesion Segmentation (ISLES) Challenge (\url{https://www.isles-challenge.org/}) significantly advances stroke image analysis by bringing together experts in neurointervention, radiology, and computer science from leading institutions. Biomedical challenges like ISLES are now considered the \textit{de facto} gold standard for algorithm comparison by the research community and have been organized multiple times, covering diverse pathologies and imaging modalities, thereby further cementing their role in advancing medical imaging technologies \cite{menze2014multimodal, sekuboyina2021verse, lalande2020emidec, antonelli2022medical, kofler2023brain, yang2023benchmarking,gomez2024apis,ma2024efficient}.

The ISLES challenge has been a recurring feature at MICCAI. The inaugural ISLES'15 focused on segmenting sub-acute ischemic stroke lesions from post-interventional MRI and acute perfusion lesions from pre-interventional MRI \cite{maier2017isles}. Subsequent challenges, ISLES'16 and ISLES'17, emphasized stroke outcome prediction by requiring the segmentation of follow-up stroke lesions from acute multimodal MR imaging and estimating patient disability scores \cite{winzeck2018isles}. ISLES'18 addressed acute stroke segmentation indirectly and cross-modally by predicting core tissue delineated in concomitant MRI from acute perfusion CT series \cite{hakim2021predicting}. The latest ISLES'22 expanded to segment acute, sub-acute, and chronic ischemic strokes in over 2000 MRI scans \cite{liew2022large, hernandez2022isles, delarosa2024}. Thus, the ISLES Challenge has garnered considerable attention over the years. 

Building on this rich history, ISLES’24 targets the segmentation of the final (post-treatment) stroke infarct, from pre-interventional acute data. The pace at which the hypoperfused tissue converts into irreversibly damaged is widely heterogeneous among patients, brain regions and time, and is driven by diverse tolerance to ischemia and differences in collateral circulation \cite{pensato2025cerebral}.
Anticipating brain tissue evolution in stroke is critical for optimizing personalized treatment decisions, as it helps determine whether (i) mechanical thrombectomy could prevent or limit infarct growth, and whether additional attempts in cases of partial recanalization have added value, (ii) the patient might benefit from additional neuroprotective therapies~\cite{munsch2024dynamic}, or (iii) the evolving infarction is likely to cause malignant brain edema, potentially requiring decompressive hemicraniectomy ~\cite{hautmann2025malignant}. Moreover, such predictions can aid in stratifying patient transfers from remote centers by estimating treatment benefit despite delays in transport to thrombectomy-capable stroke units\cite{robben2020prediction}. The prediction of tissue fate may also facilitate the inference of clinical outcomes (e.g., NIHSS or mRS scores) through biomarkers derived from stroke lesions \cite{zhang2023non, wang2021clinical}, and it may further support resource planning (for example, patients with more severe stroke lesions and predicted outcomes might require longer stays in the stroke unit before being transferred to a general neurological ward or returned to the referring hospital \cite{yang2023risk}). Given its potential to directly impact acute stroke management and long-term prognosis estimation, accurate prediction of infarct evolution remains a critical and clinically relevant challenge.

Unlike baseline imaging techniques for core estimation, which solely rely on CTP-based algorithms, ISLES'24 offers a unique, 360-degree challenge setting where most feasibly accessible clinical data are available for participants. This includes the full CT imaging suite -non-contrast CT (NCCT), CT angiography (CTA), and perfusion CT (CTP)- as well as follow-up MRI sequences, such as diffusion-weighted imaging (DWI) and apparent diffusion coefficient (ADC) maps. In addition, comprehensive clinical tabular data are provided, comprising demographics, clinical history, laboratory results, neurological scores, and outcome measures. ISLES'24 enables the integration of CT modalities both with and without contrast, supports the longitudinal assessment of brain tissue changes before and after recanalization therapy using follow-up imaging, and incorporates clinically relevant variables known to affect ischemic tissue viability. Altogether, this multimodal and temporally structured challenge design facilitates the development of more robust and generalizable segmentation models through advanced data fusion and learning strategies. In brief, the main contributions of this work are two-fold:

\begin{enumerate}
    \item The establishment of a standardized benchmark for post-treatment stroke infarct segmentation algorithms through the ISLES'24 challenge (\url{https://isles-24.grand-challenge.org/}). This includes an extensive algorithmic evaluation conducted on one of the most comprehensive and, to the best of the authors' knowledge, the first openly available longitudinal stroke dataset.
    
    \item The identification of leading solution strategies and the systematic analysis of current algorithmic limitations, along with a discussion of promising future research directions aimed at advancing infarct prediction models.
\end{enumerate}

\section{Related work}

\subsection{Final infarct prediction using perfusion CT}

Perfusion CT is the worldwide accepted imaging modality for predicting stroke lesion volumes in clinical settings. The workflow involves estimating perfusion maps using established deconvolution approaches and then deriving penumbra and core volumes through thresholding \cite{fieselmann2011deconvolution}. Although the preferred perfusion map for initial infarct (core) estimation varies among software vendors, cerebral blood flow (CBF) and cerebral blood volume maps have been widely installed. Over the past years,
various studies have proposed deep learning methods for estimating stroke lesions. For instance, de la Rosa et al. \cite{de2020differentiable} propose a deep learning-enhanced deconvolution method to estimate core tissue. Robben et al. \cite{robben2020prediction} bypass traditional deconvolution by predicting final infarct volumes using native CT Perfusion images and clinical metadata. Their approach combines diverse inputs—CTP, time, and treatment parameters, and CTP-derived vascular functions—into a DeepMedic-based model \cite{kamnitsas2017efficient}, allowing for the simulation of diverse clinical scenarios. Validation of their model over clinical data shows improved estimation of the final infarct compared to traditional deconvolution software. Building on this foundation, Amador et al. \cite{amador2021stroke, amador2022predicting} introduce a deep spatio-temporal CNN to predict final infarct from raw perfusion data. Their model combines a U-Net-like architecture with temporal convolutional networks to efficiently process 4D data. de Vries et al. present PerfU-Net \cite{de2023perfu}, a spatio-temporal model for estimating core tissue from CTP imaging data. PerfU-Net employs a U-Net-based architecture enhanced with an attention module, allowing it to capture dynamic perfusion patterns. Vision transformers have been introduced for infarct segmentation by de Vries et al. \cite{de2021transformers} and have been later combined with CNN models through hybrid methods in \cite{amador2022hybrid}. Gutierrez et al. \cite{gutierrez2024annotation} propose the prediction of follow-up CT images from CTP data using spatio-temporal models, from which the final infarct can be derived from the generated pseudo-CT images. More recently, by leveraging physics-informed neural networks, de Vries et al. \cite{de2023spatio, de2024accelerating} show promising results while using a hybrid data-driven image modeling approach.

\subsection{Beyond ASPECTS: Final infarct segmentation using non-contrast CT}
Machine and deep learning strategies for infarct segmentation have increasingly been applied to native NCCT due to its broader availability compared to advanced imaging techniques like CTA and CTP. While NCCT reveals infarcted brain tissue as hypodense areas, these changes typically become visible later than those observed with perfusion imaging. Early ischemic changes on NCCT are challenging due to low signal-to-noise ratio, subtle signals, artifacts, and confounding factors \cite{rekik2012medical}. In this context, Srivatsan et al. \cite{srivatsan2019relative} proposed an image-processing technique for segmenting early ischemic changes by creating a relative NCCT map that highlights subtle hypoattenuation differences between contralateral hemispheres. Qiu et al. \cite{qiu2020machine} trained a standard U-Net model to segment ischemia. Their model, validated over 100 independent scans, demonstrated high volumetric agreement with DWI-derived annotations. Similarly, El-Hariri et al. \cite{el2022evaluating} assessed the performance of the self-adapting nnU-Net model, showcasing high volumetric agreement with ground truth and intraclass correlations over 80\%. In 2021, Pan et al. \cite{pan2021detecting} tested a ResNet-based CNN combined with a maximum a posteriori probability model. Later on, Kuang et al. introduced EIS-Net \cite{kuang2021eis}, a multi-task learning model designed to simultaneously segment early infarcts and score the Alberta Stroke Program Early CT Score (ASPECTS) \cite{pexman2001use}. EIS-Net employs a 3D triplet convolutional neural network that incorporates original and mirrored NCCT images along with an atlas to improve feature extraction. This model excelled in both segmentation and ASPECTS scoring tasks, showing high volumetric agreement with ground-truth annotations and accurate ASPECTS ratings. Wang et al. \cite{wang2024automated}, instead, showed marginal gains of a SwinUNETR model compared to a traditional U-Net method. More recently, Kuang et al. \cite{kuang2024hybrid} leveraged circular feature interaction and bilateral difference learning in a hybrid CNN-Transformer model to segment ischemia on NCCT.

 \subsection{CTA-based final infarct prediction}
 CTA is a key imaging modality in stroke management, primarily used to assess cerebral vasculature and identify large vessel occlusions. While CTA is traditionally employed for diagnostic purposes, several studies have revealed the potential of CTA-derived features in infarct prediction~\cite{coutts2004aspects,park2019predictive}. Unlike perfusion CT, which provides direct information about blood flow and tissue viability, CTA-based approaches rely on vascular biomarkers such as collateral flow~\cite{angermaier2011ct,aoki2014collateral,elijovich2016cta} and ASPECTS~\cite{coutts2004aspects,puetz2009ct,aoki2014collateral} to estimate the final infarct size. More recently, multiphase CTA, which consists of one arterial and two venous phase acquisitions, is gaining attention and is increasingly being used for baseline core estimation \cite{shen2023prognostic, qiu2021automated}.

Sheth et al.\cite{sheth2019machine} developed an automated deep learning model called DeepSymNet to identify large vessel occlusions and the infarct core from CTA source images. Their method demonstrated an ability to determine the infarct core with an area under the ROC curve of approximately 0.88. Recently, Hokkinen et al.\cite{hokkinen2021computed} reported that a CTA-based CNN method showed a moderate correlation with final infarct volumes in the late time window for patients with large vessel occlusion who were successfully treated with endovascular therapy. Giancardo et al.~\cite{giancardo2023segmentation} further advanced the field by introducing a deep learning method that incorporates a novel weighted gradient-based approach for stroke ischemic core segmentation using image-level labeling. This approach outperforms nnU-net models trained on voxel-level data. More recently, Palsson et al.~\cite{palsson2024prediction} introduced a U-Net-based convolutional neural network for lesion prediction, augmented with a spatial and channel-wise squeeze-and-excitation block. Their study demonstrates the feasibility of predicting 24-hour follow-up infarcts using acute CTA imaging, achieving results comparable to those of CTP-based algorithms.

 \subsection{Multimodal final infarct prediction models}
 While single-modality approaches using CTP, NCCT, or CTA have shown promise in predicting final infarct volumes, there are growing research efforts on combining multiple imaging modalities for enhanced accuracy and reliability in final infarct prediction. Multimodal approaches could leverage the complementary strengths of different imaging techniques, potentially enabling a more comprehensive assessment of ischemic stroke. In this context, Wang et al.~\cite{wang2021deep} developed a hybrid multiscale CNN model that integrates NCCT, CTA, and CTA+ (acquired with an 8-second delay after CTA) to identify the infarct core and deficit. Their quantitative analysis demonstrated a high level of agreement with CTP-based definitions, highlighting the feasibility of employing commonly accessible imaging modalities (NCCT and CTA) for ischemic stroke assessment using deep learning. Other studies also explored the benefits of combining imaging data with tabular clinical information for enhanced stroke prediction. For instance, Robben et al.\cite{robben2020prediction} leverage models including images, vascular functions, and treatment information. Building on this approach, Amador et al.\cite{amador2024providing} developed a multimodal spatio-temporal model that employs cross-attention mechanisms to simultaneously fuse information from 4D CTP and clinical metadata, resulting in improved accuracy in stroke infarct prediction.

\section{Methods}
\subsection*{Dataset}
The ISLES'24 dataset includes multi-scanner and multi-center data derived from large vessel occlusion ischemic stroke cohorts. The dataset includes acute and sub-acute stroke imaging and clinical (tabular) data. Acute imaging data have been acquired at patient admission and include the diagnostic CT trilogy: NCCT, CTA, and CTP, as well as CTP-derived perfusion maps (namely CBF, cerebral blood volume (CBV), mean transit time (MTT), and time-to-maximum of the residue function (Tmax)). The follow-up imaging data were acquired 2 to 9 days later and included DWI and ADC. Moreover, the clinical data contains demographics (age, sex, etc.), patient history, admission scores (e.g., admission NIHSS), and clinical outcomes (e.g., 3-month mRS). 

The dataset is released in raw and preprocessed formats, thus allowing participants to devise algorithms with diverse degrees of freedom. For anonymization purposes, all scans are defaced based on brain and face masks obtained with TotalSegmentator \cite{wasserthal2023totalsegmentator}, and the clinical tabular data is de-identified by adding $\pm$ 5\% random noise to the laboratory variables. Preprocessing of the CTP series has been performed using the FDA-cleared clinical software ico\textbf{brain cva} \cite{de2021aifnet, de2023detecting}. The 4D CTP series are motion-corrected through image co-registration and temporally resampled at 1 frame/second. Afterwards, perfusion maps are generated using a conventional tracer-kinetics deconvolution algorithm. The CTA scans are released after cropping the images around the head, thus removing the thoracic region when available. Preprocessing of the images has been performed by linearly interpolating and registering all the imaging series to the NCCT scans. Except for the MRI scans, where affine transformations are used, all remaining images are registered following rigid transformations. Registration is performed using the Elastix \cite{klein2009elastix} and NiftyReg \cite{ourselin2002robust} toolboxes. Furthermore, MRI scans are skull-stripped using HD-BET \cite{isensee2019automated}. Lesion masks are derived from the follow-up MRI using DeepISLES \cite{delarosa2024}. Quality control and correction of the lesion masks are performed when needed by medical students (TAB, HPM) supervised by two neuroradiologists (JSK, BW) with more than 10 years of experience. Further details about the ISLES'24 data are available in the corresponding data descriptor \cite{riedel2024isles}.

The dataset (N = 248) is split into train (N = 150) and test subsets (N = 98). The train (test) set contains N = 100 (N = 50) scans from the University Hospital of Munich and N = 50 (N = 48) scans from the University Hospital of Zurich. The train subset is publicly available, while the test subset is hidden from the public, precluding participants from model overfitting strategies. The ISLES'24 data is structured and maintained following the Brain Imaging Data Structure (BIDS) guidelines \cite{gorgolewski2016brain}.
 
\begin{figure}[t!]
    \centering
    \includegraphics[width=14cm]{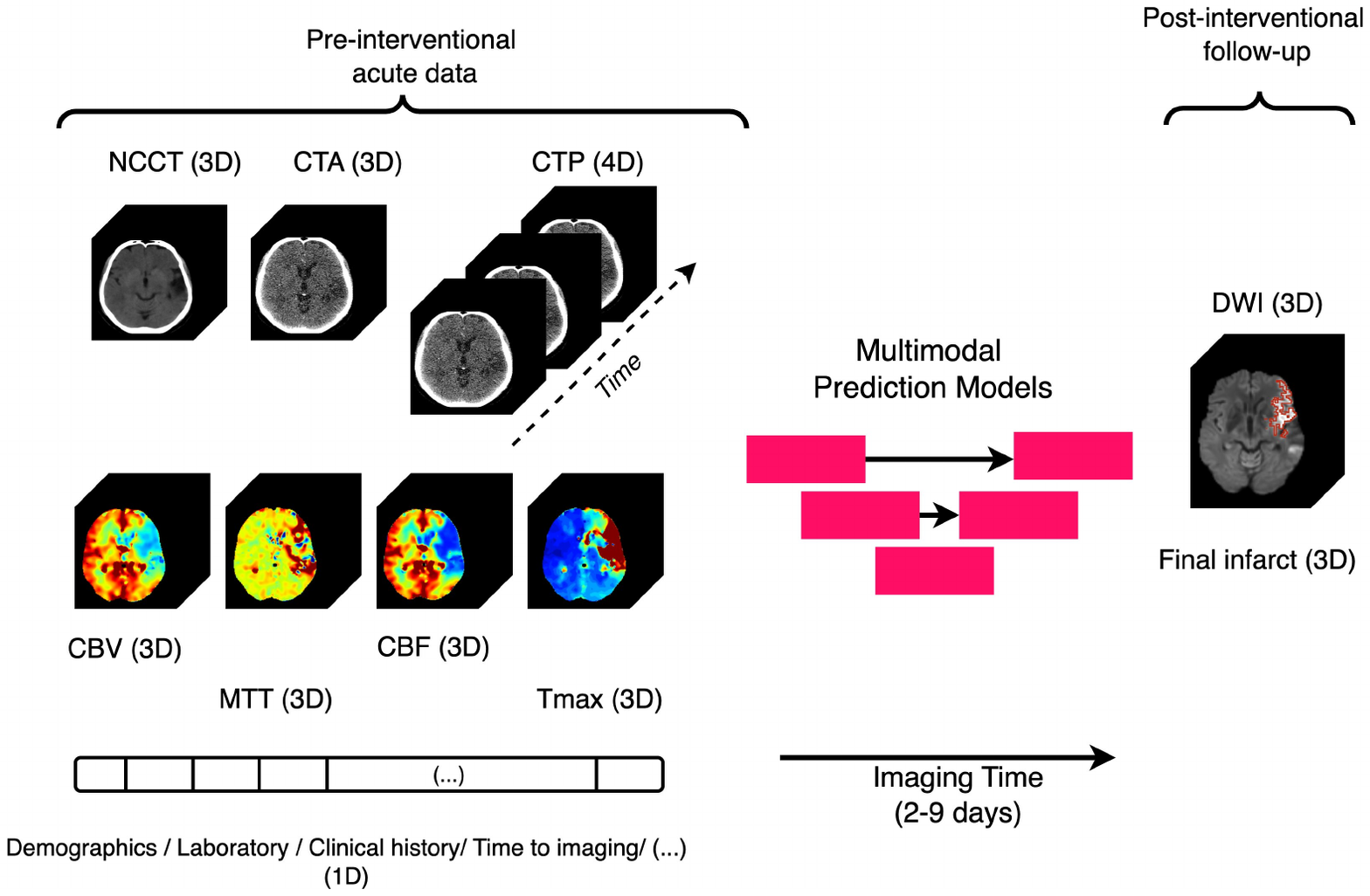}
    \caption{\textbf{Overview of the ISLES'24 task and data modalities}. Participants can exploit 1D/2D/3D/4D final infarct learning strategies. NCCT: non-contrast CT; CT: CT angiography; CTP: perfusion CT; CBV: cerebral blood volume; CBF: cerebral blood flow; MTT: mean transit time; Tmax: time-to-maximum; DWI: diffusion-weighted imaging.}
    \label{fig:isles_overview}
\end{figure}

\subsection*{Challenge set-up}
 \subsubsection*{Task}
 The ISLES'24 challenge focuses on the segmentation of final, post-treatment infarcted brain tissue based on pre-treatment imaging and clinical data. Participants are allowed to utilize multiple imaging modalities, provided either in their \textit{raw} space, in a \textit{preprocessed} (i.e., co-registered to a common image space) format, or both.
Multimodal data approaches can be leveraged, as the dataset spans multiple dimensions—from 1D tabular clinical data to 3D and 4D imaging modalities. An overview of the challenge task and the associated data modalities is provided in Figure~\ref{fig:isles_overview}.

\subsubsection*{Metrics}
The algorithmic results are evaluated by comparing the predicted lesion masks with the (manually traced) ground-truth infarct masks. The four following segmentation and task-specific, clinically relevant metrics are used: $i)$ Dice Similarity Coefficient\cite{dice1945measures} ($\mathrm{DSC= \frac{2 \cdot \text{TP}}{2 \cdot \text{TP} + \text{FP} + \text{FN}}}$, with $TP$ the true positive lesion voxels, $FP$ the false positive lesion voxels, and $FN$ the false negative lesion voxels), $ii)$ absolute volume difference ($\mathrm{AVD = |Volume_{\,predicted} - Volume_{\,ground\: truth}|}$), $iii)$ lesion-wise $\mathrm{F1_{\,score} = \frac{2 \cdot TP_{\,lesion}}{2 \cdot TP_{\,lesion} + FP_{\,lesion} + FN_{\,lesion}}}$, and $iv)$ absolute lesion count difference ($\mathrm{ALCD = |\#\, Lesions_{\,predicted} - \#\, Lesions_{\,ground\: truth}|}$). Building on previous ISLES editions \cite{delarosa2024}, where lesion-wise F1-scores were computed by considering a predicted lesion as a true positive if it overlapped with any voxel of a ground-truth lesion, the ISLES'24 challenge introduces a more stringent and objective evaluation protocol. In this edition, a predicted lesion is only considered a true positive if its spatial overlap with a ground-truth lesion exceeds a predefined threshold. This refinement is clinically motivated: accurately identifying each individual infarcted lesion is particularly relevant in cases involving multiple-territory infarctions -where lesions are distributed across distinct vascular territories- or embolic strokes, which can manifest as scattered micro-infarcts. From a clinical perspective, instance-wise segmentation can offer valuable insights into stroke etiology (e.g., multiple lesions in the acute and posterior circulation might be indicative of cardioembolism \cite{merino2010imaging}) and support more nuanced diagnostic and treatment decisions. The evaluation process consists of three main steps. First, lesion instances are extracted from both the ground-truth and predicted segmentation masks using connected component analysis. Second, predicted lesion segmentations are matched with reference annotations. This is based on computing the intersection-over-union (IoU) between all possible pairs and retaining matches that exceed an empirically selected threshold of 20\%. Finally, for each ground-truth lesion, if a matching predicted lesion is found (i.e., IoU $\geq$ 20\%), it is counted as a true positive (TP); otherwise, it is considered a false negative (FN). Predicted lesions without a corresponding ground-truth match are considered false positives (FP). Lesion-wise F1-scores are then derived from these counts. For additional details on the implementation of this instance-wise metric, we refer the reader to \cite{kofler2023panoptica}.

\subsubsection*{Ranking}
The final challenge ranking is based on the evaluation of each method's performance on an unseen test set. In this test phase, participants have no access to the input images, their corresponding outputs, or the ground truth annotations. Each algorithm is executed only once on this dataset. This setup entirely precludes any manual intervention, visual quality assessments, or case-specific overfitting strategies by the teams. The ranking is performed as in previous ISLES editions \cite{maier2017isles, winzeck2018isles, hakim2021predicting, delarosa2024} through a `rank then aggregate' strategy, using averaging as the aggregation operation. The ranking scheme is derived from the average ranking position obtained from all the individual rankings by case and by metric. First, for each scan and metric, the participants are ranked. Then, for each participant, the average rank across all the metrics for each case is computed, leading to a `per-case average rank' for each team. The final ranking is obtained by averaging these `per-case average ranks' across the entire dataset.

\subsection*{Baseline models}
Two baseline models are incorporated into the challenge framework and evaluated alongside participant submissions by predicting infarct regions on the test set. Both models utilize CTP-derived rCBF perfusion maps, computed after skull-stripping and ventricular structure removal using TotalSegmentator \cite{wasserthal2023totalsegmentator, cai2020fully}. The first baseline employs the widely used threshold of rCBF $<$ 30\%, which has been consistently validated in the literature and adopted in multiple clinical trials, showing strong agreement with infarct volume \cite{nogueira2018thrombectomy, albers2018thrombectomy, campbell2011cerebral}. The second baseline explores a broader range of rCBF thresholds, from 28\% to 46\%, selecting the threshold that maximizes the Dice coefficient on the training set. Dice was chosen as the optimization criterion to provide an alternative baseline focused on spatial overlap rather than volumetric agreement.

\subsection*{Statistical analysis}
Rankings are computed based on 1000 bootstrap samples. To compare unpaired distributions, we use the Wilcoxon rank-sum test, while paired comparisons of performance metrics across algorithms are assessed using the Wilcoxon signed-rank test, with Holm correction applied for multiple comparisons. Statistical significance is evaluated at the $\alpha = 0.05$ level. To evaluate the volumetric agreement between predicted and ground-truth lesion masks, Pearson correlation and Bland-Altman analysis \cite{bland1986statistical} are used.

\subsection*{Data and code availability}
The ISLES'24 dataset is accessible at \url{xxx}. Two Git repositories have been pre-released to facilitate team participation and metric quantification. The first repository, \url{https://github.com/ezequieldlrosa/isles24}, provides an introduction to the challenge and includes a Python notebook that demonstrates the computation of performance metrics. The second repository (\url{https://github.com/ezequieldlrosa/isles24-docker-template}) offers a template Docker algorithm, designed to streamline the creation of algorithmic Dockers. The F1-score and absolute lesion count difference metrics are computed using the \emph{panoptica} Python library \cite{kofler2023panoptica}. Connected components are identified using \emph{cc3d} \cite{cc3d}. The final challenge ranking is determined using the \emph{rankThenAggregate} function from the {\tt challengeR} \cite{wiesenfarth2021methods} package implemented in {\tt R} \cite{computing2013r}.

\section{Results}
This section presents the ISLES'24 algorithmic rankings and performance metrics obtained during the hidden test phase. For the best-performing methods, we also conduct further analysis to gain deeper insights into their performance, encompassing both quantitative and qualitative evaluations.
\subsection{Challenge overview}
The challenge took place between the 29th of May (release of the first batch of data) and the 28th of August 2024 (closing of submissions to the final test phase). A total of 453 participants registered for the challenge, from which 15 tested their algorithms in the preliminary evaluation phase, and 12 submitted a final, working solution to the test phase. After excluding participants who did not fulfill the participation requirements, 9 teams were considered for the final leaderboard.

\subsection{Description of algorithms} 
Table \ref{tab:algorithms} presents a summary of the key algorithmic aspects implemented by each team, including pre- and post-processing strategies, data inputs (modalities used in raw or NCCT-coregistered space), and primary model characteristics, such as deep learning architecture and loss functions. 

All three top-performing methods are based on nnU-Net \cite{isensee2019no}. The first- and second-ranked teams implemented similar approaches, utilizing nearly all image-data modalities except for the 4D perfusion (CTP) sequence. Unlike the winning team \emph{Kurtlab}, the second-ranked team \emph{AMC-Axolots} employed a cascaded nnU-Net approach rather than a single-model strategy. Notably, the third-ranked team, \emph{Ninjas}, relied exclusively on the 4D CTP perfusion modality and achieved strong individual-metric performance for Dice and absolute lesion count difference, ranking second and first, respectively. This outcome suggests that properly leveraging 4D CTP data can lead to competitive performance. A detailed description of these methods is provided below.

\begin{table}[!htp]
\centering
\small
\begin{tabular}{p{1.5cm} p{3cm} p{1.2cm} p{1.15cm} p{.5cm} p{1.9cm} p{1.15cm} p{2.1cm}}
\hline
\multirow{2}{*}{\textbf{Team}} & \multirow{2}{*}{\textbf{Preprocess}} & \multicolumn{3}{c}{\textbf{Input data}} & \multirow{2}{*}{\textbf{Model}} & \multirow{2}{*}{\textbf{Loss}} & \multirow{2}{*}{\textbf{Postprocess}} \\
\cline{3-5}
& & \textbf{Raw images} & \textbf{Co-reg images} & \textbf{Clin} \\
\hline
Kurtlab & Image clipping, histogram equalization, min-max normalization & \xmark  & CTA CBF CBV Tmax MTT & \xmark & nnU-Net (3D, residual encoders \cite{isensee2019automated}) & Dice + CE & \xmark \\
\hline
AMC-Axolotls & Image clipping, global normalization (NCCT and CTA), per-channel z-scoring (Tmax, MTT, CBF, CBV). Baseline core estimation (NCCT and CTA) fed as additional input to the model & NCCT & CTA CBF CBV Tmax MTT  & \xmark & nnU-Net (3D, full resolution \cite{isensee2019automated}) & Dice + CE & \xmark \\
\hline
Ninjas & Image clipping, skull stripping, foreground cropping, z-score normalization & \xmark & CTP & \xmark & nnU-Net (3D, residual encoders \cite{isensee2019automated}) & Dice + CE & \xmark \\
\hline
Dolphins & z-score normalization & \xmark & Tmax & \xmark & ResUNet (3D, 4-layer with LSTM blocks \cite{zhang2018road}) & Dice + CE & \xmark \\
\hline
ufpr\_ua & Image clipping, z-score normalization & \xmark & CTA & \xmark & nnU-Net (3D, low resolution \cite{isensee2019automated})  & Dice + CE & \xmark \\
\hline
HSK-CoreFinders & 2D isotropic resampling, cropping and padding, density windowing, z-score normalization & \xmark & CTP & \xmark & Cascade: Densenet (3D encoder, temporal pooling, 2D decoder \cite{huang2017densely}) Tmax \& CBF regression; segmentation (2D Unet) & MSE; CE + Tversky & \xmark \\
\hline
QuiiL & Image resizing and min-max normalization; Tmax thresholded at 9 s & NCCT & CTA CTP Tmax & \xmark & MoReT (3D \cite{jung20243d}) & Dice + Focal & Thresholding of outputs at 0.3 \\
\hline
MIPLAB-PrediCTP & Skull-stripping, image resampling, baseline subtraction, temporal CTP cropping, z-score normalization & \xmark & CTP & \xmark & Hybrid CNN-Transformer (2D \cite{amador2022hybrid}) & Dice & Morphological operations (opening and dilation), resampling to original space \\
\hline
MIPLAB-CTA & Image resampling skull-stripping, cropping and zero-padding, mean and max CTA projections, z-score normalization & NCCT & CTA & \cmark & Unet (2D, 4 fully connected layers \cite{ronneberger2015u}) & Dice & Transformation back to original space \\
\hline
\end{tabular}
\caption{\textbf{Summary of participating algorithms}. Co-reg: NCCT- coregistered; Clin: clinical patient data; CE: cross-entropy. LSTM: Long short-term memory. MSE: Mean squared error.}
\label{tab:algorithms}

\end{table}

\subsection{Solutions employed by the top-3 teams}

\textbf{Kurtlab (\#1)}
The algorithm employs a full-resolution 3D large residual encoder U-Net architecture based on nnU-Net (ResEnc L variant)~\cite{isensee2019no}. Inputs included CTA for structural information and vessel localization, along with CTP-derived CBF, CBV, MTT, and Tmax scans for perfusion information. Given the subtle nature of ischemic lesions in CT scans, a custom two-stage preprocessing pipeline was used to increase lesion visibility. First, skull-stripping was performed with SynthStrip~\cite{hoopes2022synthstrip}; then, all input contrasts underwent intensity windowing and histogram equalization to improve lesion visibility. Final preprocessing used the following windows: CTA (0--90 HU), CBF (0--35 mL/100g/min), CBV (0--10 mL/100g), MTT (0--20 s), and Tmax (0--7 s). When available, clinical windowing guidelines informed window settings with minor empirical adjustments. Guidelines used include: Tmax~$>$~6~s~\cite{olivot2009optimal}; CBV~$>$~2~mL/100g~\cite{wintermark2006perfusion}; CBF~$<$~17~mL/100g/min~\cite{bandera2006cbfthreshold}; and MTT~$>$~145\% of the contralateral baseline, where the 0--30 HU range captures the full range of variability~\cite{nukovic2023neuroimaging, alzahrani2023assessing, czap2021overview}. For CTA, windows were manually adjusted following~\cite{pulli2012acute}. For training, a 10-fold cross-validation strategy was used, with training executed over 1000 epochs using a batch size of 2 and a patch size of (40, 320, 320). The fold with the highest validation Dice score was selected for final submission. Standard nnU-Net preprocessing (Z-score normalization and [1,99] percentile windowing) achieved a Dice score of 21.8\% on the train hold-out set; adding custom windowing improved Dice to 31.0\%, and combining custom windowing with histogram equalization and normalization increased Dice to 31.8\%. The final submission achieved a Dice of 57\% on the hold-out set. Further details regarding training can be found in~\cite{ren2025wonisles24challengepreprocessing}. The code is available at \url{https://github.com/KurtLabUW/ISLES2024}.

\textbf{AMC-Axolotls (\#2)}
The solution aimed to predict the final infarct by leveraging baseline infarct and penumbra masks derived from NCCT and CTA scans, as well as CTP perfusion maps (CBF, CBV, MTT, and TMAX). Already co-registered images were used. All scans underwent preprocessing, including voxel value clipping to specific ranges (NCCT [0, 100] Hounsfield units, CTA [0, 200] Hounsfield units, CBF [0, 400] mL/100mL/min, CBV [0, 400] mL/100mL, MTT [0, 20] s, TMAX [0, 20] s), background value setting to zero, and skull stripping. Normalization involved global scaling for NCCT/CTA and per-channel z-scoring for the CTP-derived maps. The model employed was a nnU-Net  3D with full-resolution configuration \cite{isensee2019no}. A two-step approach was applied: models trained without perfusion maps in the first part were cascaded with models incorporating perfusion maps in the second. Training was conducted for 500 epochs using a five-fold cross-validation strategy, where the first two-thirds of the data were used for training and the remaining one-third for validation. The training process utilized the Stochastic Gradient Descent optimizer with an initial learning rate of 1e-2, weight decay of 3e-5, and momentum set to 0.99 with Nesterov acceleration. Learning rates were decayed over the epochs to ensure convergence, and no additional post-processing was applied to the model predictions. The final test set prediction was made using the best-performing model based on the Dice score. Code is available at \url{https://github.com/Mahsa0M/isles2024_docker}.

\textbf{Ninjas (\#3)}. The solution aimed to derive infarct regions by capturing contrast agent dynamics from pre-registered CTP images. First, CTP scans were clipped to values between 0 and 100 HU, followed by skull stripping using Synthstrip \cite{huo2022mapping} and foreground cropping to retain only the brain region, thereby reducing image size and enhancing training efficiency. A sequence of 20 time points centered around the peak contrast concentration was extracted to form a 20-channel input.  The model used was a 3D full-resolution nnU-NETv2 with medium residual encoder presets \cite{isensee2019no}. Training was performed with a batch size of 2 and a patch size of 28 × 256 × 192 on two NVIDIA A6000 GPUs. Z-score normalization was applied across all channels. A five-fold cross-validation approach was used, training for 100 epochs per fold with stochastic gradient descent (initial learning rate of 1e-2, weight decay of 3e-5, and Nesterov acceleration). Due to challenging time constraints, the best-performing model among the five folds, based on Dice score, was used for submission. Code is available at \url{https://github.com/jaymoz/ISLES-Challenge-2024}.

\begin{table}[t!]
\centering
{\small
\begin{tabular}{p{0.2cm}p{2.1cm}p{2.1cm}p{0.25cm}p{2.5cm}p{0.25cm}p{2cm}p{0.25cm}p{2.2cm}p{0.25cm}}
\hline
\multicolumn{1}{c}{\textbf{FR}} & \multicolumn{1}{c}{\textbf{Algorithm}} & \multicolumn{2}{c}{\textbf{Dice}} & \multicolumn{2}{c}{\textbf{AVD}} & \multicolumn{2}{c}{\textbf{F1 (\%)}} & \multicolumn{2}{c}{\textbf{ALCD}} \\ \hline
& \multicolumn{1}{c}{} & \multicolumn{1}{c}{\textbf{Stats}} & \multicolumn{1}{c}{\textbf{R}} & \multicolumn{1}{c}{\textbf{Stats}} & \multicolumn{1}{c}{\textbf{R}} & \multicolumn{1}{c}{\textbf{Stats}} & \multicolumn{1}{c}{\textbf{R}} & \multicolumn{1}{c}{\textbf{Stats}} & \multicolumn{1}{c}{\textbf{R}} \\ 
\hline
1 & Kurtlab   & \textbf{0.285} (0.213) & 1 & \textbf{21.23} (37.22) & 3  & 14.40 (21.20) & 2 & 7.18 (7.67) & 3 \\ \hline
2 & AMC-Axolotls & 0.263 (0.247) & 3 & 21.31 (35.23) & 4 & \textbf{14.94} (25.12) & 1 & 7.66 (7.94) & 5 \\ \hline
3 & Ninjas & 0.255 (0.191) & 2 & 26.29 (39.73) & 7 & 9.92 (13.46) & 5 & \textbf{5.98} (6.46) & 1 \\ \hline
4 & Dolphins & 0.178 (0.209) & 8 & 22.33 (34.88) & 5 & 9.31 (20.13) & 4 & 7.63 (7.99) & 4 \\ \hline
5 & ufpr\_ua & 0.199 (0.227) & 7 & 22.72 (39.44) & 6 & 11.97 (23.57) & 3 & 7.91 (8.05) & 7 \\ \hline
6 & HSK-CoreFinders & 0.201 (0.160) & 4 & 36.79 (29.88) & 9 & 3.01 (5.92) & 6 & 8.29 (6.66) & 6 \\ \hline
6 & rCBF $<$ 0.3$^{*}$ & 0.163 (0.147) & 5 & 21.27 (33.39) & 1  & 0.46 (1.28) & 8 & 139.12 (101.36) & 9 \\ \hline
6 & rCBF $<$ 0.44$^{*}$ & 0.191 (0.157) & 5 & 27.50 (29.84) & 1  & 0.47 (0.97) & 8  & 194.34 (112.30) & 9  \\\hline 
9 & QuiiL & 0.068 (0.090) & 11 & 168.06 (112.87) & 11 & 1.12 (5.06) & 10 & 7.11 (7.54) & 2 \\ \hline
9 & MIPLAB-PrediCTP & 0.139 (0.158) & 9 & 77.38 (68.23) & 10 & 1.79 (3.47) & 7 & 20.02 (12.72) & 8 \\ \hline
11 & MIPLAB-CTA & 0.073 (0.095) & 10 & 27.51 (38.80) & 8 & 0.11 (0.45) & 11 & 125.34 (39.41) & 11 \\ \hline 
\end{tabular}
}
\caption{\textbf{Summary of performance metrics for each team, presented as mean (standard deviation) and rank}. FR denotes the final team rank, while R represents the metric-specific rank. The best mean values for each metric are highlighted in bold. The baseline models are indicated with $^{*}$.}
\label{tab:team_metrics}
\end{table}

\subsection{Leaderboard and Algorithmic Performance}
\begin{figure}[t!]
    \centering
    \includegraphics[width=\textwidth]{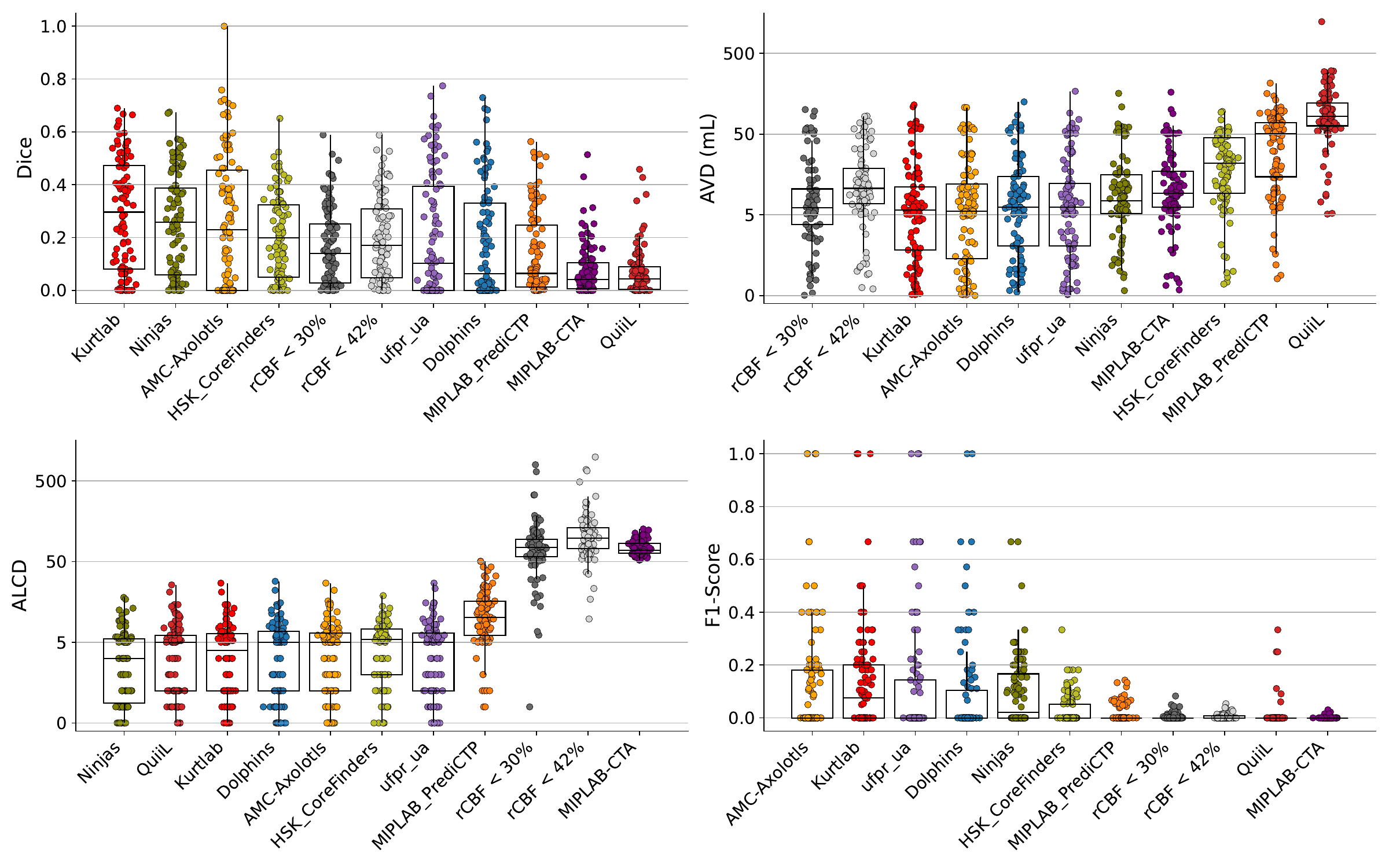}
    \caption{\textbf{Boxplots by performance metrics in the test phase}. For each metric, algorithms are ranked from best (left) to worst (right). For visualization purposes, a non-linear scale is used for AVD and ALCD plots. AVD: Absolute volume difference. ALCD: Absolute lesion count difference.}
    \label{fig:boxplots}
\end{figure}

The test set leaderboard is presented in Table~\ref{tab:team_metrics}, and the distribution of the evaluation metrics is shown in Figure~\ref{fig:boxplots}. The optimal threshold identified for the baseline model is rCBF $<$ 0.44. The winning team of the challenge, \emph{Kurtlab}, achieved the highest individual rank in Dice, second place in F1-score, and third place in both absolute volume difference and lesion count difference. The second-ranked team, \emph{AMC-Axolotls}, led the F1-score ranking, while the third-ranked team, \emph{Ninjas}, achieved the best performance in absolute lesion count difference. 

Notably, the baseline models based on rCBF achieved the top rank in absolute volume difference.

\begin{figure}[t!]
    \centering
    \includegraphics[width=13cm]{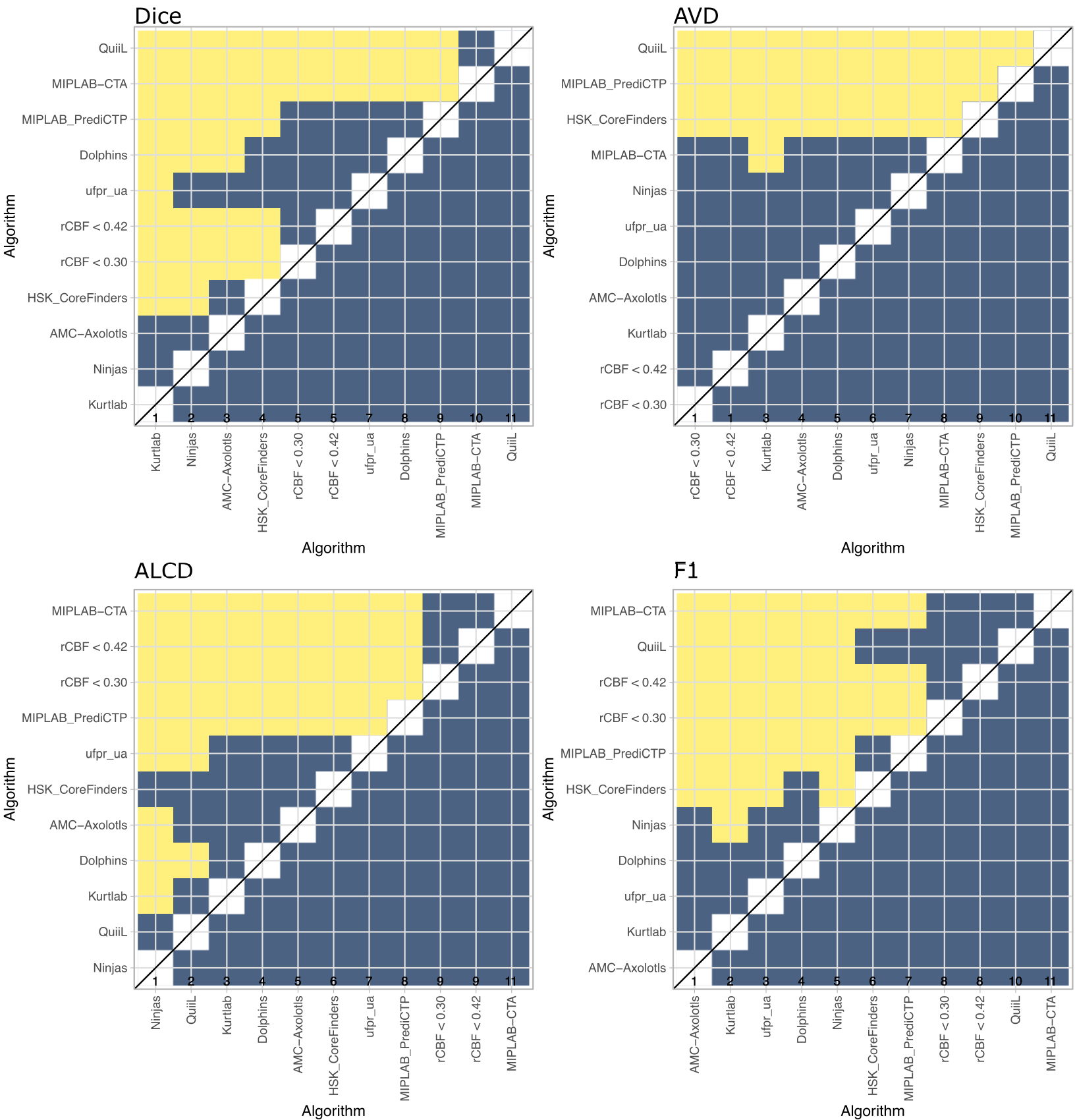}
    \caption{\textbf{Significance maps across algorithms for each metric based on a 1000-bootstrap experiment}. Statistical significance is assessed using the Wilcoxon signed-rank test with Holm correction for multiple comparisons. ALCD: Absolute lesion count difference. AVD: Absolute volume difference.}
    \label{fig:sign_maps}
\end{figure}

Significance maps across algorithms and metrics are presented in Figure~\ref{fig:sign_maps}. No statistically significant differences are observed in Dice scores among the top three ranked solutions. Furthermore, no statistically significant differences are found in any metric between the two top-ranked teams, \emph{Kurtlab} and \emph{AMC-Axolotls}, indicating comparable algorithmic performance. The third-ranked team, \emph{Ninjas}, demonstrated statistically significant improvements in lesion count difference compared to nearly all other methods. Although the baseline models (rCBF~$<$~30\% and rCBF~$<$~44\%) led the absolute volume difference ranking, they did not show statistically significant improvements in this metric when compared to the top three teams.

\subsection{Ranking stability}
\begin{figure}[hbtp!]
    \centering
    \includegraphics[width=\textwidth]{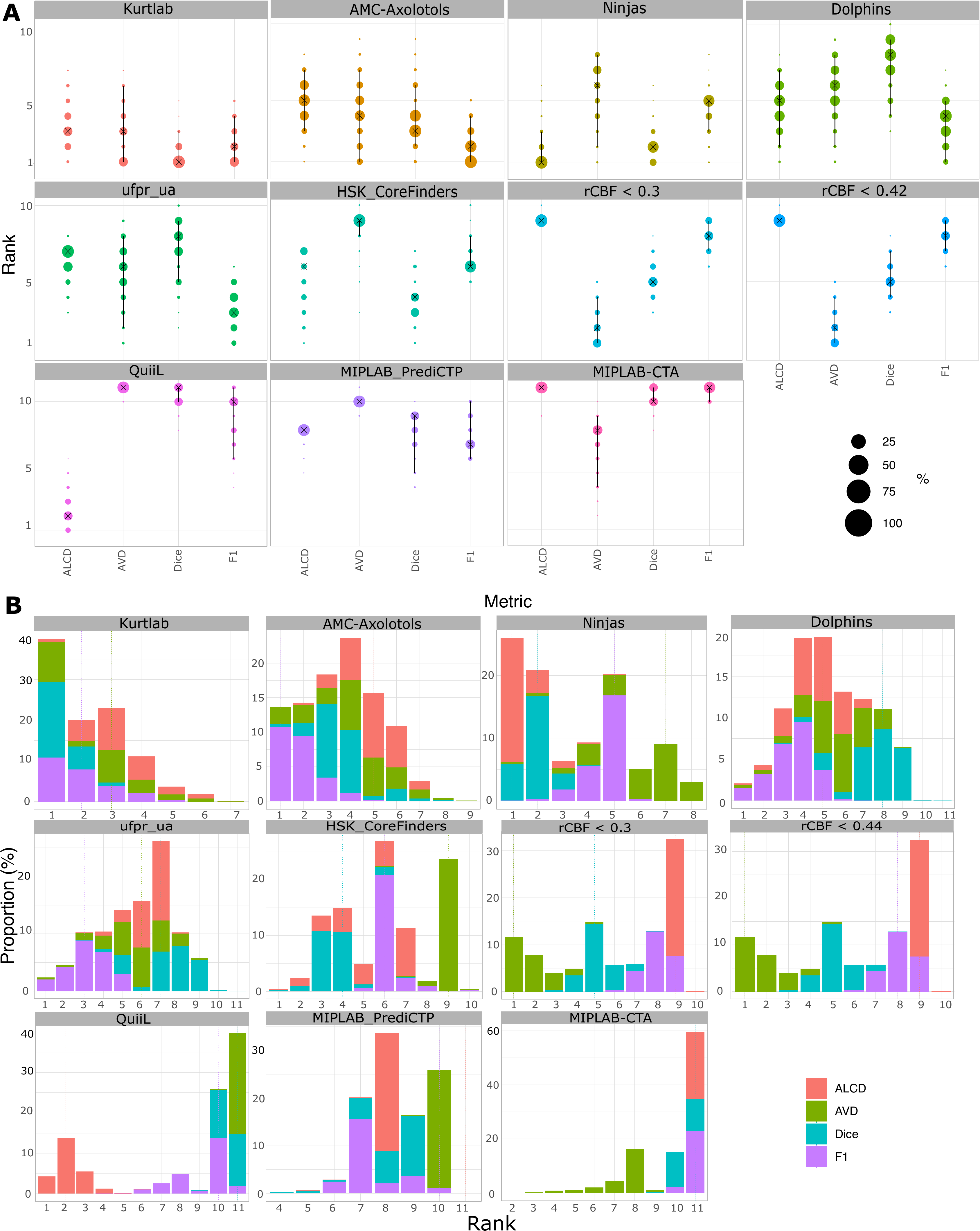}
    \caption{\textbf{Test-set ranking stability based on a 1000-bootstrap analysis}. \textbf{A:} Blob plot illustrating the distribution of ranks for each algorithm and evaluation metric. The size of each blob reflects the proportion of bootstrap samples in which a team achieved a given rank. \textbf{B:} Stacked frequency plots of observed ranks per team and metric, with vertical lines indicating the final rank achieved on the full test set. Methods with more consistent high (low) performance exhibit left- (right-) skewed distributions. ALCD: absolute lesion count difference. AVD: absolute volume difference.
}
    \label{fig:stab}
\end{figure}

To better understand leaderboard robustness and algorithmic variability on the test set, we performed a bootstrap analysis with 1,000 resamples across all evaluation metrics. Figure~\ref{fig:stab} summarizes the resulting ranking stability. Blob plots confirm that the challenge winner, \emph{Kurtlab}, consistently ranks first in Dice and maintains competitive rankings in lesion-wise F1 scores and absolute volume difference. In line with previous findings, \emph{Ninjas} achieves the highest ranking for estimating the number of lesions per scan, while perfusion-based baselines rank best for final infarct volume estimation.

Examining the stacked frequency plots, \emph{Kurtlab} exhibits the most right-skewed and top-rank-centered distribution, further supporting its status as the leading solution in the challenge. In addition, \emph{Kurtlab} demonstrates notable robustness, with all metric rankings falling within the top seven positions. The second-ranked team, \emph{AMC-Axolotls}, also shows a right-skewed distribution, although slightly shifted toward lower ranks. Notably, \emph{Ninjas} achieves the highest proportion of top-ranked cases in absolute lesion count estimation. The perfusion-based baselines demonstrate consistent performance in terms of absolute volume difference, with all bootstrap-derived ranks confined to the top four positions for this metric.

\subsection{Infarct size impacts performance}
\begin{figure}[t!]
    \centering
    \includegraphics[width=\textwidth]{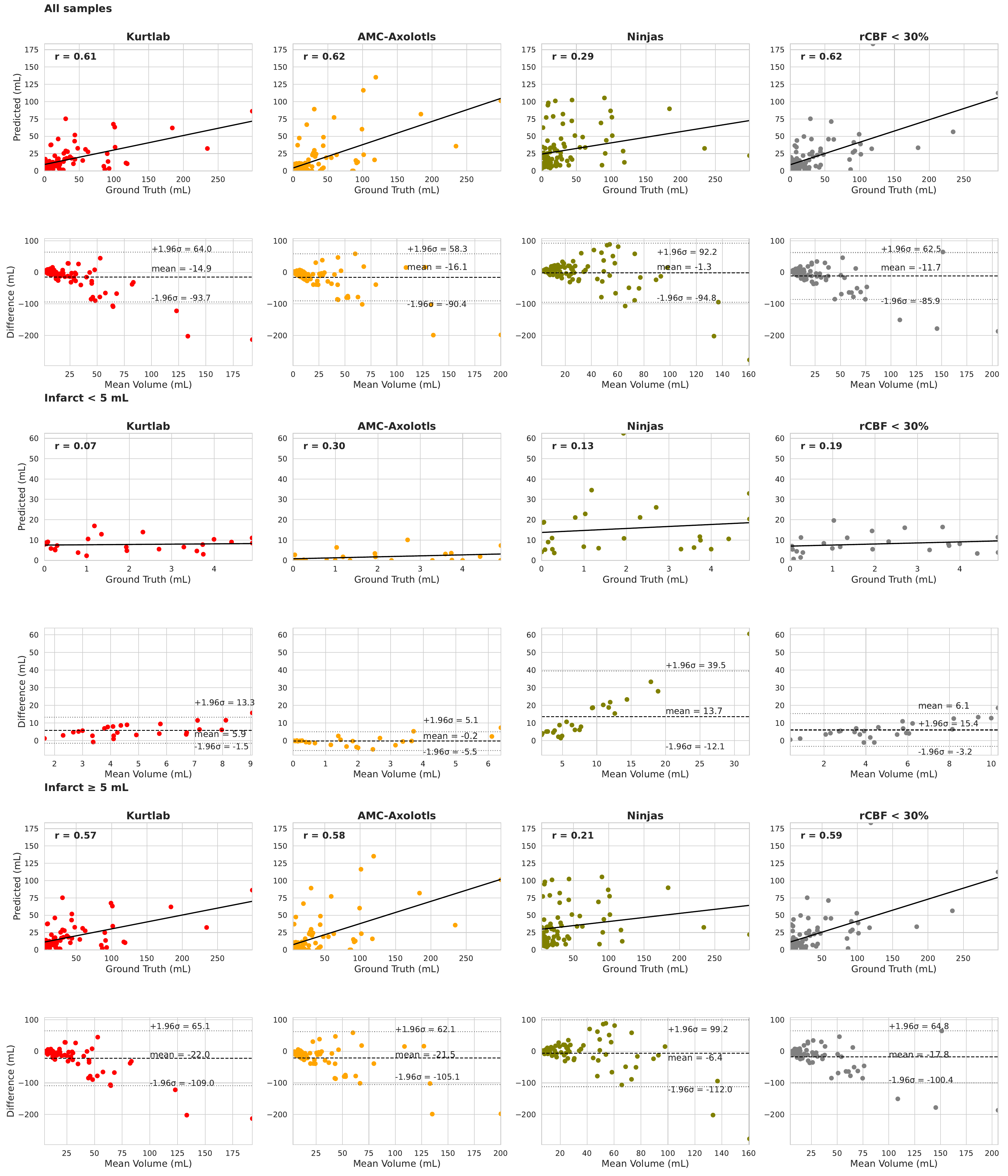}
    \caption{\textbf{Volumetric agreement for the top-3 methods and the clinical baseline (rCBF $<$ 0.3) stratified by stroke infarct size}. Scatter (first row) and Bland-Altman (second row) plots are shown. Difference values in Bland Altman plots are obtained by subtracting the ground truth values from the predicted ones. r: Pearson correlation coefficient.  $\sigma$: Standard deviation.}
    \label{fig:vol_agreement}
\end{figure}

The dataset used in this study includes a diverse range of infarct types and sizes \cite{riedel2024isles}. Identifying small lesion instances (e.g., emboli) is significantly more challenging than segmenting large or extensive stroke infarcts \cite{delarosa2024}. To assess model performance, we divide the test set into two groups: cases with a total infarct volume smaller than 5 ml (including small infarctions, emboli, or ischemic lesions introduced during recanalization treatment) and those with a total infarct volume of 5 ml or greater. Figure \ref{fig:vol_agreement} illustrates the volumetric agreement for the top three algorithms, stratified by lesion size.  

Considering the full test set (\(N=98\)), the highest volumetric agreement in terms of Pearson correlation is achieved by both \emph{AMC-Axolotls} and the perfusion baseline model using rCBF~$<$~30\%, with each reaching a Pearson \(r = 0.62\). \emph{Ninjas} exhibit the lowest volumetric bias, with a mean volumetric difference of 1.29~mL, though at the expense of greater volumetric variability compared to other top-performing algorithms. Across methods, there is a general tendency to underestimate the final infarct volume.

When focusing on cases with small infarcts (\(\text{infarct} < 5~\text{mL}\), \(N=25\)), all methods show limited volumetric agreement (Pearson \(r < 0.4\)). The best performance in this subset is again observed for \emph{AMC-Axolotls}, which achieves a Pearson correlation of \(r = 0.3\) and a mean volumetric bias of -0.2~mL. Despite this, the prediction of small infarcts remains particularly challenging for all teams, as reflected by generally low Dice scores. The highest Dice scores for small lesions are obtained by \emph{Kurtlab} and \emph{AMC-Axolotls}, with Dice values of 0.14~$\pm$~0.19 and 0.14~$\pm$~0.26, respectively. In contrast, \emph{Ninjas} and the rCBF~$<$~30\% baseline report substantially lower performance, with Dice scores of 0.09~$\pm$~0.12 and 0.04~$\pm$~0.06, respectively, underscoring the difficulty of accurately delineating subtle ischemic regions.

For infarcts equal to or larger than 5\,mL ($N=73$), the baseline model based on rCBF\,$<$\,30\% and the \emph{AMC-Axolotls} team exhibit the strongest volumetric agreement with the ground truth, achieving Pearson correlation coefficients of $r = 0.59$ and $r = 0.58$, respectively. \emph{Ninjas} show the lowest volumetric bias, with a mean volume difference of $-6.4$\,mL, albeit with greater variability than other methods. In terms of spatial overlap, \emph{Kurtlab} achieves the highest Dice score ($0.34 \pm 0.20$), followed by \emph{Ninjas} ($0.31 \pm 0.18$) and \emph{AMC-Axolotls} ($0.30 \pm 0.23$). The rCBF-based baseline obtained a Dice score of $0.20 \pm 0.14$. While Dice scores improve with increasing lesion volume, considerable performance variability persists.

\subsection{Multi-center generalization}
\begin{figure}[ht]
    \centering
    \includegraphics[width=\textwidth]{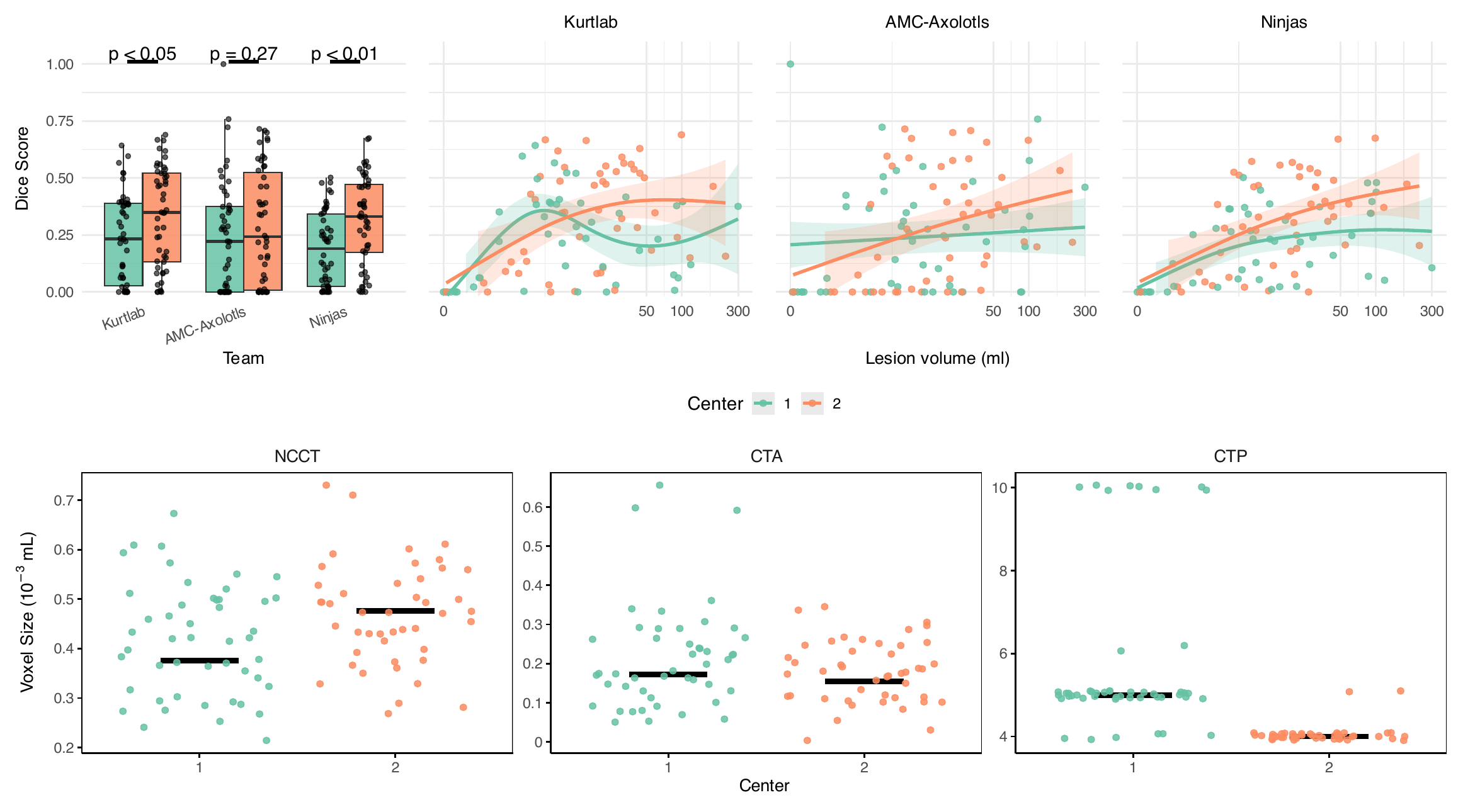}
    \caption{\textbf{Multicenter generalization.} 
\textbf{Top:} Performance of the top-3 algorithms by acquisition center (boxplots) and across lesion sizes (scatter plots). For larger lesions, most algorithms tend to perform better on cases from Center~\#2 compared to Center~\#1. 
\textbf{Bottom:} Voxel resolution of test-set cases grouped by center and modality. CTP acquisitions from Center \#2 include scans with higher resolution compared to those from Center \#1. In the scatter plots above (below), the x-axis (y-axis) is displayed using a non-linear scale to better visualize the wide range of values. \textit{p}-values are obtained with Mann--Whitney U tests.
}
    \label{fig:multicenter}
\end{figure}

We evaluate the consistency of algorithm performance across data from the two acquisition centers. Figure~\ref{fig:multicenter}A displays Dice scores for the top-performing algorithms, revealing statistically significant superior performance on cases from Center~\#2 compared to Center~\#1. This finding is unexpected, as the training data includes twice as many cases from Center~\#1, and any performance gap would have been anticipated in the opposite direction. To investigate whether lesion size differences could explain this discrepancy, Figure~\ref{fig:multicenter}A also presents scatter plots of Dice scores versus lesion volume, stratified by center. However, for all top-3 teams, performance on Center~\#2 remains consistently higher even at comparable lesion sizes, suggesting that the observed performance gap is not attributable to differences in infarct volume.

 We investigate whether differences in center-specific image acquisition protocols may be associated with the observed performance gap. Figure~\ref{fig:multicenter}B shows the distribution of voxel sizes (scan resolution) across the three CT-based modalities (NCCT, CTA, and CTP) for each center. While Center~\#1 provides higher resolution NCCT images, its CTP acquisitions exhibit consistently a lower resolution compared to Center~\#2. CTA resolutions appear comparable between centers.

Two hypotheses may explain these findings. First, team \emph{Ninjas}, which ranked among the top three, relied exclusively on CTP data. This may account for their much significant better performance on Center~\#2 cases, where higher-quality CTP acquisitions were available. Methods based solely on CTP may be more sensitive to differences in acquisition protocols, resulting in decreased generalizability. Second, team \emph{AMC-Axolotls}, the only top-3 team whose performance differences between centers were not statistically significant, employed a two-stage cascaded approach, with an initial model segmenting infarcts without using CTP. This strategy likely leverages the superior NCCT resolution available in Center~\#1, reducing dependency on lower-quality CTP data and contributing to more balanced generalization across centers. These findings might also provide a plausible insight into why CTP-only methods underperformed relative to more comprehensive, multimodal approaches in the challenge.


\subsection{Performance by tissue region}
\begin{figure}[b!]
    \centering
    \includegraphics[width=\textwidth]{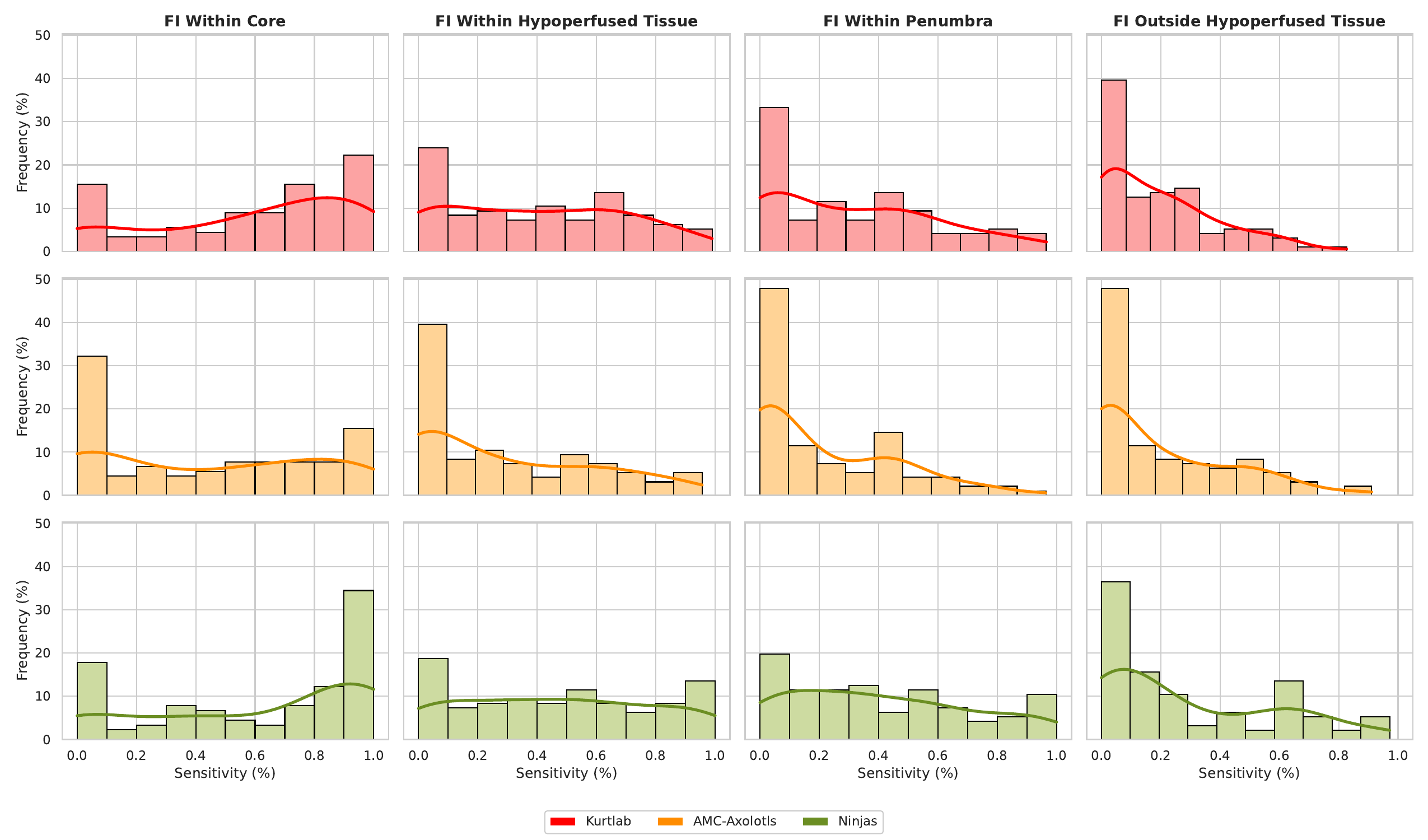}
\caption{\textbf{Algorithmic sensitivity by perfusion-defined tissue regions in the test set for the top-performing solutions.} Better (worse) performance is reflected by right- (left-) skewed sensitivity distributions. FI: Final infarction.}

    \label{fig:se_tissue}
\end{figure}

To assess how algorithms perform across different brain regions affected by ischemia, we analyzed the sensitivity of the top three teams in detecting final infarction within perfusion-defined tissue regions. These regions were derived from acute perfusion imaging and follow standard clinical definitions \cite{nogueira2018thrombectomy, albers2018thrombectomy}: (i) \textit{core} tissue, defined as regions with rCBF~$<$~30\%; (ii) \textit{hypoperfused} tissue, defined as regions with Tmax~$>$~6\,s; (iii) \textit{penumbra} tissue, corresponding to the hypoperfused region excluding the core; and (iv) \textit{non-hypoperfused} tissue, defined as regions with Tmax~$\leq$~6\,s. Figure~\ref{fig:se_tissue} illustrates the detection sensitivity of each algorithm within these regions.

Better (worse) algorithmic performance is reflected in right- (left-) skewed sensitivity distributions. Sensitivity progressively decreases from the \textit{core} region -representing the most evidently infarcted tissue- to the \textit{penumbra}, and further to regions \textit{outside} the hypoperfused territory. The latter represents the most challenging detection scenario, which may occur in cases without perfusion deficits on acute imaging, or in more complex situations such as post-stroke edema, which can increase intracranial pressure, compress brain structures, and lead to midline shifts~\cite{gu2022cerebral}. The second-ranked team, \emph{AMC-Axolotls}, shows overall lower sensitivity performance, reflected by more left-skewed distributions, compared to the other leading teams. Interestingly, the third-ranked team, \emph{Ninjas}, achieves higher sensitivity within the \emph{penumbra} and even outside the \emph{hypoperfused} tissue region when compared to both competitors. Team \emph{Ninjas} based their solution on CTP imaging, which corresponds to the modality used to define the tissue regions in this analysis. Therefore, this alignment may explain their competitive performance, particularly within the \emph{core} region. However, \emph{Ninjas} also outperform the other teams in detecting infarction beyond the hypoperfused tissue, suggesting that their method captures ischemic injury not readily identified through conventional CTP-derived parameter maps.

\subsection{Qualitative results}
\begin{figure}[b!]
    \centering
    \includegraphics[width=\textwidth]{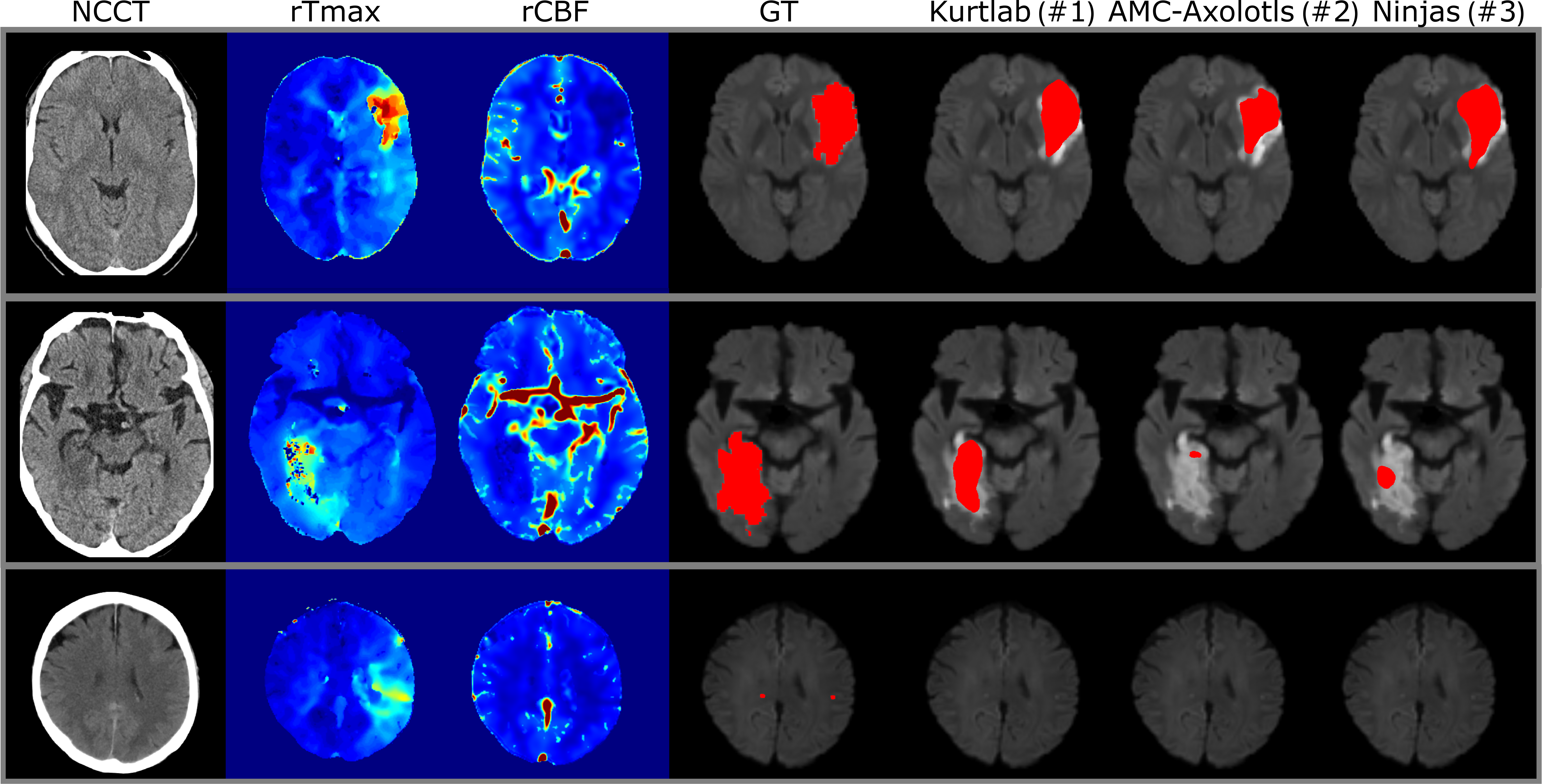}
    \caption{\textbf{Qualitative infarct prediction results.} Each row shows a different test-set scan corresponding to the best (top), median (middle), and worst (bottom) performance cases, with average Dice scores of 0.68, 0.28, and 0.00, respectively. GT: ground truth; rTmax: relative (healthy-tissue normalized) Tmax.}
    \label{fig:quality}
\end{figure}

Example segmentation results from the top three teams are presented in Figure \ref{fig:quality}. The scan with the highest Dice score demonstrates distinct perfusion deficits in Tmax and CBF, which align well with the follow-up final infarct. All methods successfully capture the lesion to some extent. The median Dice case (average Dice across methods: 0.28) highlights a notably good performance by the challenge-winning team, \emph{Kurtlab}. In the last row, a case of punctiform embolic stroke is presented, characterized by hypoperfused tissue visible on the Tmax map but lacking clear evidence of irreversible damage on CBF. None of the methods detects the embolic lesions, yielding a Dice score of zero. However, no method erroneously predicts the hypoperfused region as infarcted tissue, thereby correctly estimating the tissue fate. In total, there are five test-set scans yielding Dice scores of zero for the three teams: two consist of small single-lesion infarctions ($<$ 5~mL), two include punctiform lesions associated with scattered microinfarcts, and the remaining case exhibits no final infarction.

\section{Discussion}
Predicting final infarct lesions from pre-interventional acute data is inherently complex due to several factors. While acute CT imaging is widely accessible and routinely used in clinical practice, it is inherently less sensitive than the gold-standard DWI and is often affected by artifacts (e.g., streak artifacts) and a low signal-to-noise ratio. These technical limitations pose significant challenges in identifying small infarcts, such as scattered micro-infarcts, which are essential for comprehensively assessing stroke extent and understanding its underlying etiology. Beyond imaging limitations, the dynamic and heterogeneous nature of stroke progression introduces further complexity. Distinct infarct growth phenotypes—ranging from slow to fast progressors—have been identified, influenced by patient-specific pathophysiological variables, particularly the status of leptomeningeal collateral circulation, which sustains residual perfusion and limits infarct expansion \cite{pensato2025cerebral, munsch2024dynamic, vagal2018collateral}. Notably, infarct growth may continue despite successful mechanical recanalization \cite{regenhardt2021infarct}, and additional scattered lesions may be induced during the recanalization procedure itself, compounding the prediction task. Moreover, technical challenges related to data preprocessing—such as image registration and interpolation across differing image spaces, resolutions, and modalities (e.g., acute CT and follow-up MRI)—further hinder accurate lesion segmentation. Altogether, this multifaceted problem underscores the necessity for advanced modeling strategies and meticulous data handling in stroke lesion prediction tasks.

The ISLES'24 challenge provides a unique framework for identifying algorithmic solutions capable of predicting final stroke infarction by leveraging the full spectrum of imaging and clinical data available at patient admission. This comprehensive setting facilitates a 360-degree understanding of both patient condition and stroke pathophysiology. To date, and to the best of our knowledge, no other study in the field has provided such a comprehensive dataset coupled with systematic algorithmic benchmarking. Our results highlight that the well-established nnU-Net architecture \cite{isensee2019no} consistently yields the best performance for this task, with the top three teams on the leaderboard adopting variations of this model. Notably, the first- and second-ranked teams integrated all available imaging modalities—excluding 4D CTP—and outperformed nnU-Net-based models relying solely on single modalities, such as teams utilizing only CTP or CTA data. Given that current clinical workflows for final infarct estimation predominantly rely on NCCT or CTP-derived perfusion maps, our findings underscore that multimodal data integration represents a key direction for advancing infarct prediction in future clinical practice. Notably, the perfusion-based baseline models using rCBF achieved infarct volume estimates comparable to those of the best-performing deep learning methods. However, deep learning approaches outperformed in voxel-wise metrics, demonstrating superior spatial localization and identification of infarction. These results highlight the ability of deep learning to enhance lesion detection and overlap at the voxel level while maintaining volumetric agreement with traditional perfusion-based models. Developing deep learning strategies that meaningfully surpass the volumetric accuracy of rCBF-based baselines remains an open challenge.


\subsection{Open questions and methodological opportunities}
Our challenge benchmark highlights a significant research gap in effectively exploiting the interplay between imaging data and clinical tabular information. Although one team incorporated clinically relevant variables -such as stroke severity scores at admission, patient demographics, clinical history, and critical time metrics- there was no demonstrated tangible improvement in segmentation performance through the integration of these features. A standard concatenation-based fusion strategy was used, which proved insufficient to capture the complex relationships between imaging and clinical variables. Whether more sophisticated feature fusion strategies, such as attention-based mechanisms, cross- and multi-modal techniques like CLIP~\cite{radford2021learning}, or advanced tabular data architectures such as TabFPN~\cite{hollmann2025accurate}, can enhance segmentation performance, remains an open question. While prior studies have successfully integrated clinical and imaging data for functional outcome prediction tasks~\cite{jo2023combining, borsos2024predicting, liu2023functional, hatami2022cnn}, their application in infarct growth prediction models is still scarce~\cite{robben2020prediction, amador2021stroke}. Further algorithmic investigation is required to determine whether the incorporation of clinical tabular data can substantively improve final infarct prediction models.

More robust and refined models capable of effectively exploiting the 4D CTP modality are still needed. Perfusion imaging is widely recognized as one of the most informative imaging techniques for assessing cerebral hypoperfusion and identifying regions of irreversibly damaged tissue. However, our challenge benchmark reveals two key findings: $(i)$ models utilizing standard deconvolution-derived perfusion maps consistently outperform those trained directly on raw 4D CTP series, and $(ii)$ models that extract spatial features from the 4D CTP data achieve superior performance compared to those attempting to leverage the full spatio-temporal nature of the modality. Specifically, the top-performing model utilizing raw 4D CTP data (team \emph{Ninjas}) selected a subset of meaningful CTP timepoints and applied a spatial convolutional nnU-Net architecture, whereas models designed to explicitly capture temporally-encoded contrast propagation dynamics consistently underperformed. These observations emphasize the current limitations in modeling the spatio-temporal characteristics of perfusion imaging. We thus advocate for the development of advanced spatio-temporal models -such as those proposed in \cite{amador2021stroke, amador2022hybrid, amador2022predicting, de2024accelerating, de2023spatio, de2023perfu}- to more effectively capture perfusion dynamics, and encourage their validation on comprehensive datasets like ISLES'24.

Our findings suggest that the inclusion of NCCT may play a key role not only in lesion detection but also in enhancing model generalization across acquisition centers. It is well established in the literature that NCCT can reliably identify infarcted regions presenting as hypodense tissue at admission, with numerous studies demonstrating the feasibility of infarct segmentation directly from NCCT \cite{srivatsan2019relative, qiu2020machine, el2022evaluating, pan2021detecting, kuang2021eis, wang2024automated, kuang2024hybrid} or through ASPECTS-based scoring \cite{pexman2001use}. However, our results indicate that NCCT may also provide a practical advantage in mitigating variability associated with CTP acquisition protocols, which often differ substantially across institutions due to scanner vendors, contrast injection parameters, and post-processing pipelines. As a more standardized and higher-resolution modality, NCCT offers a stable input source that may reduce model sensitivity to domain shifts in CTP data. These findings support the potential value of multimodal approaches that incorporate NCCT alongside CTP and CTA, suggesting that robust infarct segmentation models can benefit not only from richer input information but also from the complementary and generalizable characteristics of diverse imaging modalities.

\subsection{Limitations}
Our study is limited by the dataset size and population diversity, as the cohort exclusively comprises European patients. In contrast to other stroke datasets that may offer larger cohorts but are typically limited to cross-sectional data or single imaging modalities, ISLES'24 emphasizes the provision of a comprehensive longitudinal, multimodal patient dataset to enable the development of more holistic algorithmic solutions. However, curating such a dataset is inherently more challenging: it is constrained by missing data modalities, requires substantial annotation effort, and involves extensive qualitative controls. Despite these efforts, external validation of the proposed algorithms on larger and more diverse datasets remains essential to confirm generalizability.

Furthermore, the inclusion criteria of the ISLES'24 dataset were restricted to patients with favorable recanalization outcomes (i.e., thrombolysis in cerebral infarction -TICI- scores 2B and 3). Consequently, this study does not capture brain tissue evolution in patients with poor recanalization outcomes or those who did not undergo thrombectomy. Future work may benefit from incorporating such patient populations and developing models that account for these clinical scenarios, as performed in \cite{debs2021impact}. 

Finally, the ISLES'24 dataset does not include the full digital subtraction angiography image series, but provides expert-rated recanalization scores based on the TICI scale. Future datasets and benchmarking efforts could be further enhanced by incorporating complete digital subtraction angiography sequences, thereby enabling more comprehensive modeling of stroke pathophysiology and treatment response.

\section*{Acknowledgement}
SW is supported by the Swiss National Science Foundation (Grant No. 310030\_200703), the UZH Clinical Research Priority Program (CRPP) Stroke, and the Swiss Heart Foundation. EdlR and BM are supported by the Helmut Horten Foundation. EOR is supported by the TUM KKF Clinician Scientist Program. KA, AW, NDF, AB, EU, and MW are supported by the Heart and Stroke Foundation of Canada. HB is supported by the Koetser Foundation and the Young Talents in Clinical Research program of the SAMS and of the G. \& J. Bangerter-Rhyner Foundation.

\end{document}